\documentclass[review]{elsarticle}

\usepackage[pdftex,
        colorlinks=true,
        urlcolor=darkblue,               
        anchorcolor=darkblue,
        filecolor=green,                   
        linkcolor=darkblue,              
        menucolor=darkblue,
        citecolor=darkblue,
        pagebackref,
        backref=false,
        pdfpagemode=true,
        bookmarks=true,
        bookmarksopen=true,
        ]{hyperref}

\usepackage[
uncertainty-mode = separate, 
exponent-product = \cdot, 
]{siunitx}
\DeclareSIUnit{\bohrmagneton}{\ensuremath{\mu_\mathrm{B}}}
\DeclareSIUnit{\Coatom}{\text{Co atom}}
\usepackage{booktabs}

\usepackage{upgreek} 
\usepackage[utf8]{inputenc}
\usepackage[british]{babel}

\journal{Journal of Magnetism and Magnetic Materials}

\biboptions{numbers,sort&compress}

\begin{document}

\begin{frontmatter}

\title{Influence of thickness on magnetic properties of RF-sputtered amorphous CoNbZr thin films}

\author[1]{Indujan Sivanesarajah\corref{cor1}}
 \ead{sivanesa@outlook.de}
\author[2]{Leon Abelmann}
\author[1]{Uwe Hartmann}

\cortext[cor1]{Corresponding author}
\address[1]{Institute of Experimental Physics, Saarland University, D-66041 Saarbr\"{u}cken, Germany}
\address[2]{Department of Microelectronics, Delft University of Technology, 2600 AA Delft, The Netherlands}

\begin{abstract}
Amorphous sputtered Co-based thin films are widely used as soft magnetic materials in applications such as sensors, inductors and magnetic flux concentrators. The magnetic properties of these films can be controlled by deposition parameters like film thickness, argon pressure, deposition rate and others. In this study, we present a detailed investigation of the magnetic properties of RF-sputtered Co$_{91}$Nb$_7$Zr$_2$ films with thicknesses ranging from \qty{52}{nm} to \qty{1040}{nm}. These amorphous films exhibit an average saturation magnetisation of \qty{1.01(4)}{MA/m}. As the film thickness increases, there is a significant decrease in coercivity, remanent-to-saturation magnetisation ratio $M_\mathrm{r}/M_\mathrm{s}$, and maximum permeability. The change in macroscopic magnetic properties is also reflected by the domain structure. At a thickness of \qty{52}{nm}, the remanent domain state shows irregular domains, while films thicknesses above \qty{208}{nm} exhibit flux-closure domain structures instead. The thickness-dependent modifications are attributed to the transition between N\'eel and Bloch type domain walls, which is expected to occur at approximately \qty{84}{nm}. 
\end{abstract}


\end{frontmatter}


\section{Introduction}

Amorphous alloys have gained substantial prominence as soft magnetic materials \cite{Shimada1982, Sakakima1983, Herzer2013}. These materials find versatile applications, ranging from magnetic sensors \cite{Meydan1994, Treutler2001, Moron2015} and inductors \cite{Korenivski2000, Rylko2009,Hsiang2022} to motors \cite{Hasegawa2004, Chai2020, Enomoto2023}. Of particular interest is CoNbZr, a specific variant of amorphous Co-based alloys, known for its high susceptibility, large saturation magnetisation, and low coercive field \cite{deWit1987,Zuberek1992}. These properties make CoNbZr an ideal choice for integrated magnetic flux concentrators, establishing it as a material of paramount importance in this area \cite{Marinho2011, He2018, Maspero2021}. 

CoNbZr films are typically prepared by sputter deposition. However, the magnetic properties can be considerably influenced by details of the deposition process. For instance, different argon pressures can induce changes in the magnetic properties due to residual stress and morphology alterations, leading to the emergence of perpendicular magnetic anisotropy \cite{Peng2013}. Furthermore, Takahashi et al. revealed intriguing insights into how a compressed magnetic field magnetron sputtering technique affects key magnetic parameters \cite{Takahashi1989}. 

Of particular interest is the pivotal role that the film thickness plays in determining the magnetic performance of CoNbZr films. Observations involve an inverse relationship between film thickness and coercive field \cite{Li2007, Cao2012, Takahashi1989} and an increasing anisotropy field with increasing film thickness \cite{Katada2000}. An increasing film thickness also leads to an intriguing in-plane uniaxial magnetic anisotropy \cite{Cao2012}. 

However, despite extensive research in the field, a comprehensive understanding of how fine film thickness influences the magnetic properties in samples prepared through RF-magnetron sputtering is not yet existing. In particular, the intricate relationship between the film thickness and soft magnetic properties, such as permeability has not been thoroughly explored. This present study conducts an examination of the magnetic properties of RF-sputtered CoNbZr films at varying film thickness. 

\section{Materials and methods}

This section describes the fabrication of CoNbZr thin films and the methods used for their structural and magnetic characterisation.

\subsection{Sample preparation}

CoNbZr films were deposited on Si - SiO$_2$ substrates of \qtyproduct{8x8}{mm}, mounted on a rotating holder using RF-magnetron sputtering. The target composition Co$_{85}$Nb$_{12}$Zr$_3$ (at. \%) was chosen due to the nearly zero magnetostriction ($\lambda_\mathrm{s} < 10^{-6}$) \cite{Yamaguchi2015}. The target-to-substrate distance was set to \qty{12}{cm}. After reaching a base pressure below \qty{1e-8}{mbar}, the films were deposited at an argon pressure of \qty{1.7e-3}{mbar} at \qty{100}{W}, while maintaining a deposition rate of \qty{4.9(2)}{nm/min}. By adjusting the sputtering time, films with thicknesses between \qty{52}{nm} and \qty{1040}{nm} were fabricated.

For magnetic domain structure analysis via magnetic force microscopy \allowbreak (MFM), CoNbZr films were patterned into \qtyproduct{20x20}{\um} squares using a focused ion beam (\textsc{FEI Helios NanoLab 600}) at an acceleration voltage of \qty{30}{kV}. Initially, an outer frame was patterned to coarsely remove material at a beam current of \qty{21}{nA}, followed by patterning of the final structures at a current of \qty{6.5}{nA}. No ion beam imaging was performed in the patterned area to prevent detrimental effects of Ga ions on the magnetic properties of the structures \cite{Fassbender2008}.

\subsection{Microstructural characterisation}
X-ray diffraction (XRD) patterns of the investigated samples were recorded at room temperature using a D8-A25-Advance diffractometer (Bruker, Karlsruhe, Germany) in Bragg-Brentano $\theta - \theta$ geometry, with a goniometer radius of \qty{280}{nm} and Cu $\mathrm{K}_\alpha$ radiation ($\lambda$ = \qty{154.0596}{pm}). A \qty{12}{\um} Ni foil served as a $\mathrm{K}_\beta$ filter and a variable divergence slit was positioned at the primary beam side. A \textsc{lynxeye} detector with \num{192} channels was employed on the secondary beam side. Experiments were conducted in a $2\theta$ range of \qtyrange{30}{60}{\degree} with a step size of \qty{0.013}{\degree} and a total scan period of \qty{4}{hours}.

Electron diffraction patterns were acquired using a \textsc{JEOL 2011} transmission electron microscope (TEM), equipped with a LaB$_6$ cathode and operated at \qty{200}{kV}. TEM slices were prepared by milling a \qtyproduct{15x5}{\um} area from the continuous film using a focused ion beam (FIB) followed by in situ lift-out with a micromanipulator \cite{Tomus2013}. The elemental composition of the TEM slices was determined using an energy-dispersive X-ray spectroscopy (EDX) system.

\subsection{Magnetic characterisation}

Magnetic measurements were performed on the as deposited films using a \textsc{DMS} Model 10 Vector Vibrating Sample Magnetometer (VSM) to measure both in-plane and out-of-plane magnetic hysteresis loops. The lateral sample size was \qtyproduct{8x8}{mm}. The system was calibrated using a \qty{0.3}{mm} thick Ni foil with the same area and a mass of \qty{156}{mg}, assuming a magnetic moment of \qty{8.61}{mAm^2} based on the specific magnetisation of Ni (\qty{55.1}{Am^2/kg}) \cite{Crangle1971}. The diamagnetic contributions from the sample holder and Si substrate were subtracted as a linear background signal of \qty{0.5}{\micro Am^2/T}, obtained by fitting the high-field branches of the in-plane hysteresis loops.

For MFM analysis, a Bruker Multimode 8 atomic force microscope (AFM) was used. The AFM head was positioned between two electromagnetic coils capable of applying a field $B$ of up to \qty{80}{mT}. Cantilevers (Olympus OMCL-AC240TS) coated with \qty{40}{nm} Co$_{85}$Cr$_{15}$ were used to detect the stray magnetic fields of the samples \cite{Babcock1994}.
\section{Results and discussion}

This section presents the structural and magnetic properties of the CoNbZr thin films, followed by a discussion of key hypotheses based on the observed trends.

\subsection{Microstructural properties}

\begin{figure}[t!]
\centering
\includegraphics[width=\linewidth]{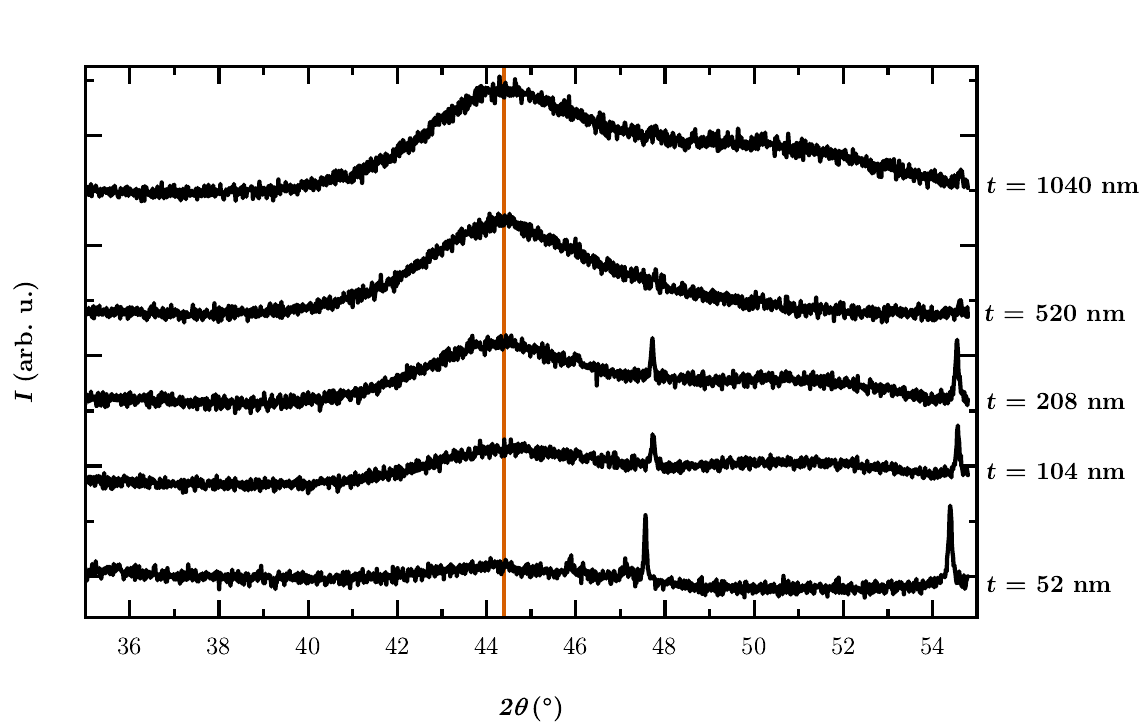} 
\caption{XRD patterns of sputtered CoNbZr films with varying thicknesses $t$. The $y$-axis represents intensity $I$ and the $x$-axis represents the diffraction angle $2\vartheta$. The broad peaks around \qty{44.40(6)}{\degree} (red line) indicate the presence of an amorphous phase, confirming that the CoNbZr films retain their amorphous structure as thickness increases. Two distinct Bragg reflections at \qty{47.67(9)}{\degree} and \qty{54.50(9)}{\degree} are attributed to the crystalline structure of the underlying (\num{100}) Si substrate.}
\label{fig:XRD}
\end{figure}

\begin{table}[!b]
\centering
\caption{Peak centre positions and FWHM of Co extracted from the XRD pattern.}
\sisetup{
table-alignment-mode = format,
table-number-alignment = center,
separate-uncertainty = true
}
\begin{tabular} {@{\extracolsep{\fill}}
S[table-format=-4.0+-2.0]
S[table-format=-2.2+-1.2]
S[table-format=-1.2+-1.2]
@{ }}
{$t$} & {Peak centre} & {FWHM} \\
{nm} & {\qty{}{\degree}} & {\qty{}{\degree}} \\
\hline
1040(40) & 44.47(2) & 4.30(4)\\
520(20) & 44.40(1) & 4.17(3)\\
208(8) & 44.38(2) & 4.01(5)\\
104(4) & 44.44(5) & 3.75(10)\\
52(2) & 44.32(8) & 2.96(5) \\
\hline
 & 44.40(6) & 3.84(53) \\
\end{tabular}
\label{XRD_parameters_Co}
\end{table}

\begin{table}[t]
\centering
\caption{Peak centre positions and FWHM of Si extracted from the XRD pattern.}
\begin{tabular} {@{\extracolsep{\fill}}
S S S S S S
@{ }}
{$t$} & {Peak centre} & {FWHM} & {Peak centre} & {FWHM} \\
{nm} & {\qty{}{\degree}} & {\qty{}{\degree}} & {\qty{}{\degree}} & {\qty{}{\degree}} \\
\hline
\num{208(8)} & \num{47.72(1)}& \num{0.07(1)} & \num{54.55(1)} & \num{0.09(1)} \\
\num{104(4)} & \num{47.73(1)} & \num{0.09(1)} & \num{54.57(1)} & \num{0.09(1)} \\
\num{52(2)} & \num{47.56(1)} & \num{0.07(1)} & \num{54.40(1)} & \num{0.10(1)} \\
\hline
 & \num{47.67(9)} & \num{0.08(1)} & \num{54.50(9)} & \num{0.09(1)}\\
\end{tabular}
\label{XRD_parameters_Si}
\end{table}

Figure \ref{fig:XRD} shows the X-ray diffraction (XRD) patterns of CoNbZr films with varying thicknesses. The $y$-axis represents intensity $I$ and the $x$-axis represents the diffraction angle $2\vartheta$. In the $2\vartheta$ range of interest, broad, low-intensity peaks centred at approximately \qty{44.40(6)}{\degree} were observed, with a full-width at half-maximum (FWHM) of around \qty{3.84(53)}{\degree} (see Tab.\ref{XRD_parameters_Co}). The broadening of these peaks is due to the scattering of X-rays in multiple directions, caused by the lack of a regular atomic lattice and a statistically dominating specific range of interatomic distances \cite{Cullity2014}. This broad peak is characteristic of an amorphous state, indicating the absence of long-range crystalline order in the CoNbZr films.

In addition to the broad peak, two distinct Bragg reflections were observed at \qty{47.67(9)}{\degree} and \qty{54.50(9)}{\degree} (see Tab. \ref{XRD_parameters_Si}). These reflections are attributed to the crystalline structure of the underlying (\num{100}) Si substrate, likely induced by incident W $\mathrm{L}_\beta$ radiation originating from the X-ray source, which is equipped with a tungsten filament. These Si-related reflections are distinct from the broad peak associated with the amorphous CoNbZr films.

\begin{figure}[t!]
\centering
\includegraphics[width=0.5\linewidth]{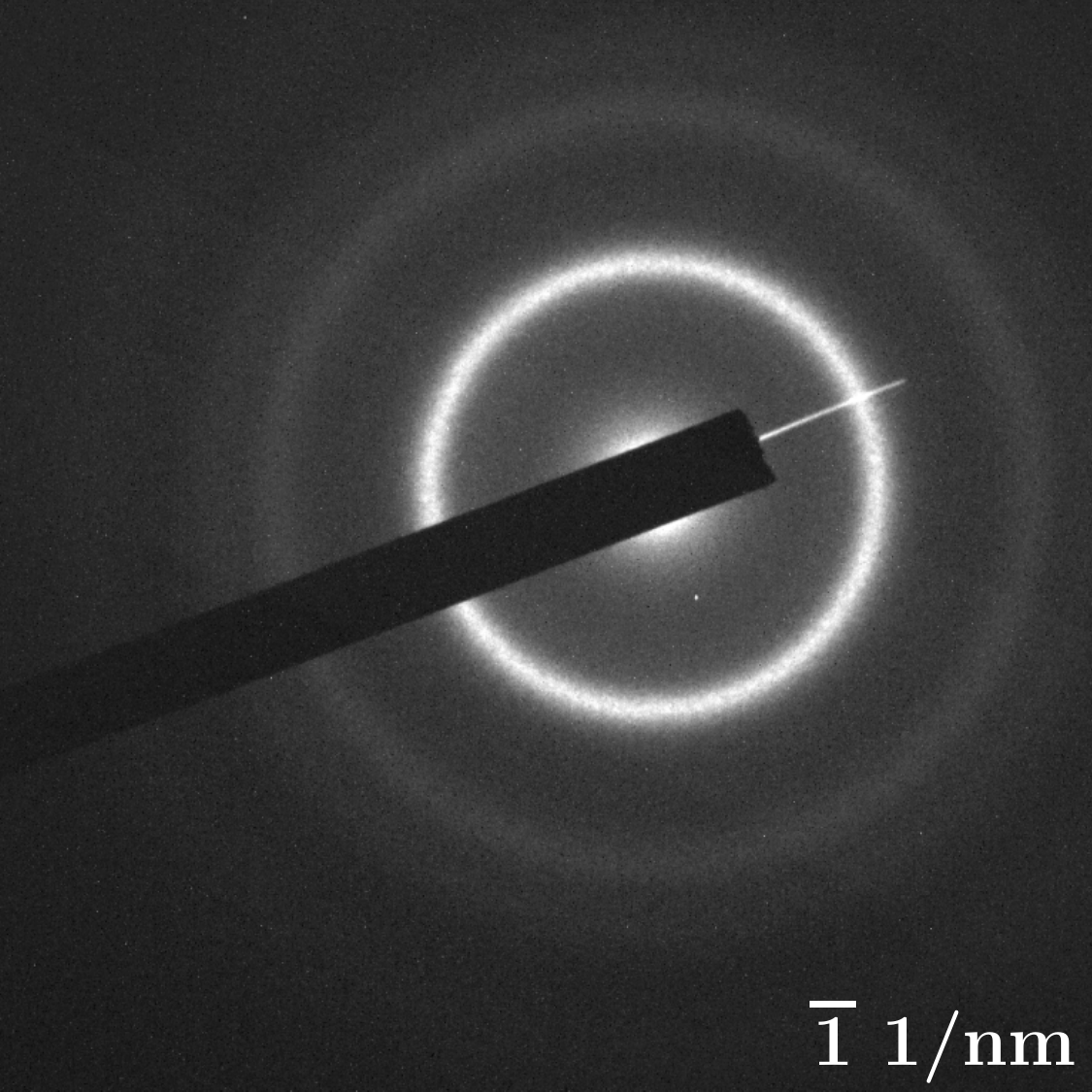} 
\caption{Electron diffraction pattern of a sputtered CoNbZr film with a thickness of \qty{208}{nm}. The presence of diffuse rings is indicative of an amorphous state.}
\label{fig:TEM}
\end{figure}

While the XRD data suggest an amorphous structure, they do not allow to differentiate between nanocrystalline and amorphous states. To address this, electron diffraction measurements were conducted. Figure \ref{fig:TEM} displays the electron diffraction pattern of a CoNbZr film with a thickness of \qty{208}{nm}. The presence of diffuse rings, rather than discrete diffraction spots, is consistent with an amorphous structure, corroborating the XRD findings. 

EDX analysis demonstrated that all sputtered films consistently exhibit a stoichiometry of Co$_{91 \pm 1}$Nb$_{7 \pm 1}$Zr$_{2 \pm 1}$, showing a slight deviation from the target composition. This variation is likely due to differences in the deposition rates, which can be influenced by factors such as sputtering pressure, target-substrate distance, and the significant disparities in atomic mass and radius \cite{Neidhardt2008}.

\subsection{Magnetic properties}

\begin{figure}[t!]
\centering
\includegraphics[width=\linewidth]{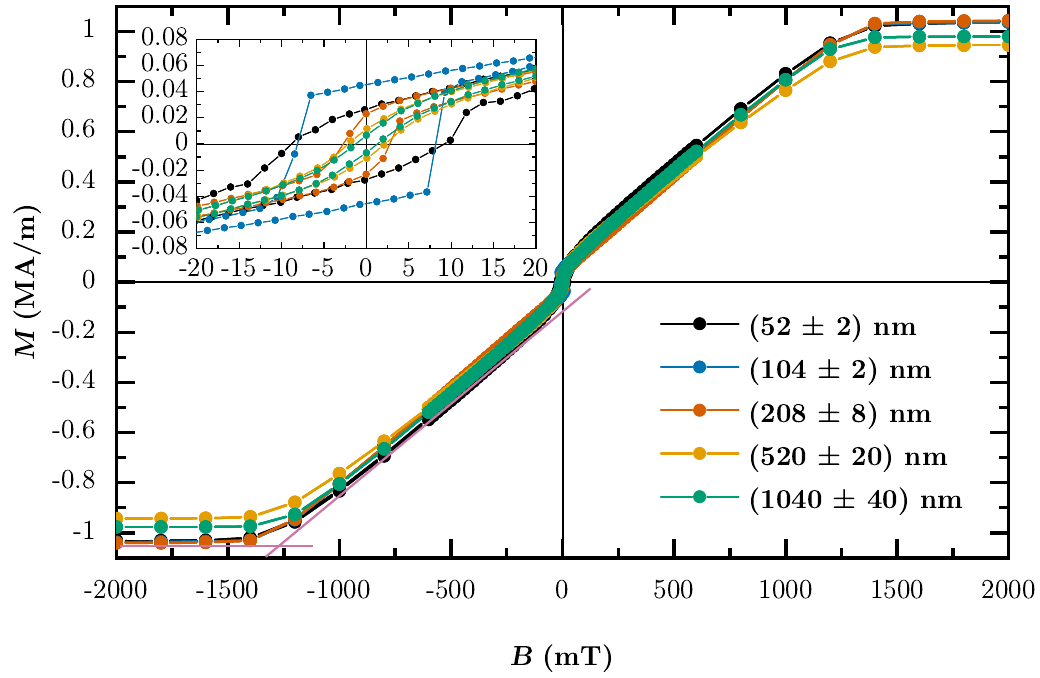} 
\caption{Out-of-plane hysteresis loops of the magnetisation $M$ for amorphous CoNbZr films of varying thicknesses, measured using VSM at a field sweep of $B =\pm$\qty{2000}{mT}. The pink lines indicate how the saturation field was estimated (see Fig. \ref{fig:Ms}).}
\label{fig:Hys_perp}
\end{figure}

\begin{figure}[t!]
\centering
\includegraphics[width=\linewidth]{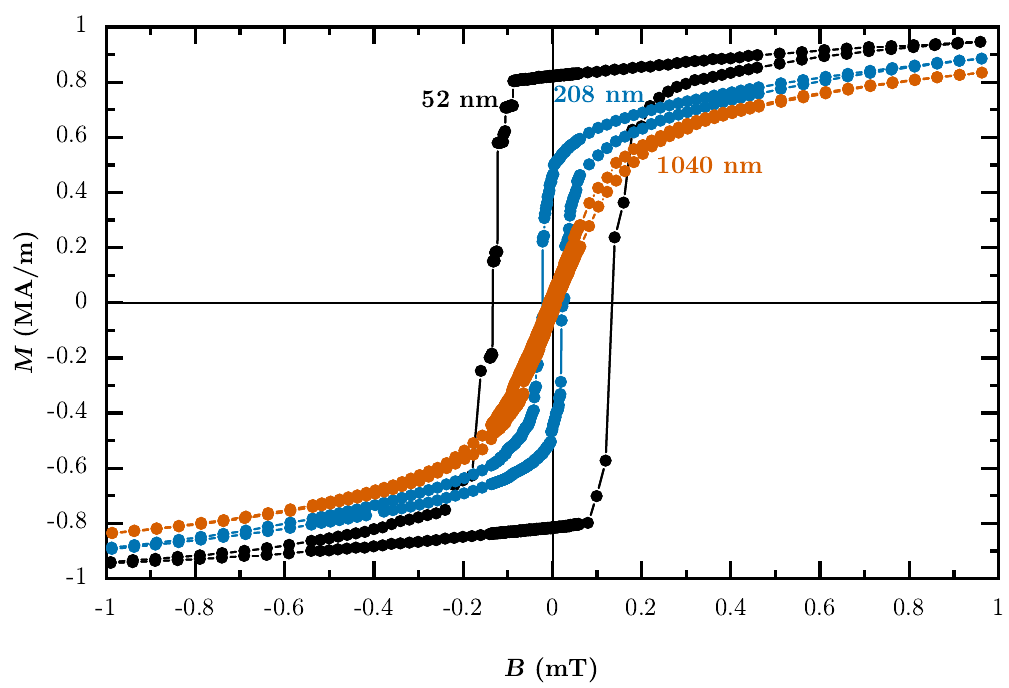} 
\caption{In-plane hysteresis loops of the magnetisation $M$ for a field sweep of $B =\pm$\qty{1}{mT} for amorphous CoNbZr films measured by VSM at different thicknesses. As the thickness increases, the hysteresis loops transition from a square to a smoother shape. Barkhausen jumps are also observed in the thinnest film.}
\label{fig:Hys_para}
\end{figure}

Figure \ref{fig:Hys_perp} presents the out-of-plane hysteresis loops of amorphous CoNbZr films with varying thicknesses, measured using VSM at a maximum field sweep of $B=\pm$\qty{2000}{mT}. Due to a slight inevitable misalignment of approximately \qty{1}{\degree} to \qty{2}{\degree} between the film plane normal and the external field direction, a small in-plane magnetisation component is observed, affecting the low-field region of the loops. Between \qty{-1400}{mT} and \qty{-20}{mT}, the magnetisation decreases proportionally with the external field as it rotates toward the plane of the film. At \qty{-20}{mT}, the in-plane component reaches \qty{-20}{mT}$\cdot\sin(2^\circ)\approx$\qty{-0.7}{mT}, nearly sufficient to saturate the films in-plane (see Fig. \ref{fig:Hys_para}). Consequently, the magnetisation rotates further within the film plane, leading to a small magnetisation anomaly between \qty{-20}{mT} and \qty{20}{mT} (see inset in Fig. \ref{fig:Hys_perp}). At \qty{20}{mT}, the films are saturated in-plane, and further increases in the magnetic field cause the magnetisation to rotate out-of-plane, eventually saturating at \qty{1400}{mT}.

\begin{figure}[t!]
\centering
\includegraphics[width=\linewidth]{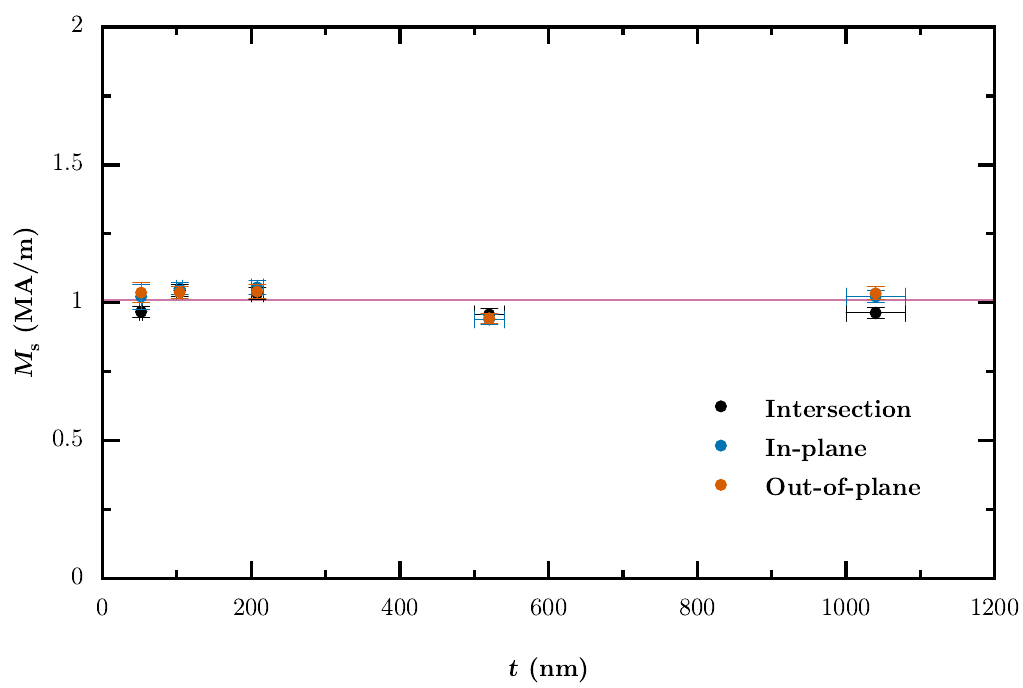} 
\caption{Derived saturation magnetisations for various film thicknesses from the out-of-plane (black and red) and in-plane (blue) loops. The analysis shows that the saturation magnetisation is independent of film thickness (indicated by the pink line), with a mean value of \qty{1.01(4)}{MA/m}.}
\label{fig:Ms}
\end{figure}

The saturation magnetisation $M_\mathrm{s}$ was estimated from the saturation magnetic moments obtained from both the in-plane and out-of-plane loops (see Figs. \ref{fig:Hys_perp}, \ref{fig:Hys_para}), considering a total film area of \qty{64}{mm^2} and the thicknesses measured by TEM. The uncertainties in $M_\mathrm{s}$ arise from inaccuracies in the saturation moment measurement, alignment errors, and the precision of the film dimensions. Additionally, the slopes of the hard-axis loops can be used to estimate $M_\mathrm{s}$ by calculating their intersections with the horizontal lines extrapolated from the saturation region (as indicated by the pink lines in Fig. \ref{fig:Hys_perp}). The latter assumes that in an ideal thin film, the out-of-plane hysteresis curve is a straight line where the saturation field equals the saturation magnetisation ($M_\mathrm{s} = N_z B_\mathrm{s}/\mu_0$, with $N_z = 1$, the out-of-plane demagnetisation factor). Unlike the previous method, this approach allows the determination of $M_\mathrm{s}$ without the film’s geometric dimensions.

Both methods yield consistent estimates for $M_\mathrm{s}$, as shown in Fig. \ref{fig:Ms}. The analysis demonstrates that the saturation magnetisation is independent of film thickness, with an average value of \qty{1.01(4)}{MA/m}. The reported error represents the standard deviation of $M_\mathrm{s}$.

\begin{table}[t!]
\centering
\caption{Comparison of saturation magnetisations $M_\mathrm{s}$ with reported values and estimated magnetic moment per Co atom $n_\mathrm{B}$. The latter is much lower than for bulk cobalt.}
\label{Ms_comparison}
\begin{tabular}{
 @{\extracolsep{\fill}}
l@{}
l@{}
S[table-format=0.2]
S[table-format=0.2]
c@{}}
\toprule
{Composition} & \multicolumn{1}{c}{$t$} & {$M_\mathrm{s}$} & {$n_\mathrm{B}$} & {Reference}\\
\multicolumn{1}{c}{(at.$\%$)} &
\multicolumn{1}{c}{(nm)} &
\multicolumn{1}{c}{(MA/m)} &
\multicolumn{1}{c}{$(\mu_\text{B}/\text{Co atom})$} &
\\
\midrule
Pure Co & Nanowire & 1.44 & 1.71 &\cite{Sun2005}\\
Co$_{96}$Zr$_{4}$ & 150 - 2000$^*$ & 1.19 & 1.36 &\cite{Shimada1982}\\ 
Co$_{91}$Nb$_7$Zr$_2$ & 52 - 1040 & 1.01 & 1.09 & This work\\
Co$_{90}$Zr$_{10}$ &150 - 2000$^*$& 0.88 & 0.94 &\cite{Shimada1982}\\
Co$_{90}$Nb$_5$Zr$_5$ &100 - 300$^*$ & 0.86 & 0.92 &\cite{Zuberek1992}\\ 
Co$_{88}$Nb$_8$Zr$_4$ & 1000 & 1.11 & 1.16 &\cite{Takahashi1989}\\
Co$_{87.3}$Nb$_{8.8}$Zr$_{3.9}$ & 1000 & 0.80 & 0.83 &\cite{deWit1987}\\
Co$_{85}$Nb$_{10}$Zr$_5$ & 2000 & 0.88 & 0.89 &\cite{Sakakima1983}\\
\bottomrule 
\end{tabular}
\begin{minipage}{10cm}
\footnotesize{$^*$Exact thickness $t$ at which measurements were performed is not reported.}
\end{minipage}
\end{table}

When comparing the derived saturation magnetisation with previous studies (see Tab. \ref{Ms_comparison}), the present value is higher than those reported for similar CoNbZr compositions. For instance, prior studies have reported $M_\mathrm{s}$ values of approximately \qty{0.80}{MA/m} for Co$_{87.3}$Nb$_{8.8}$Zr$_{3.9}$ \cite{deWit1987}, \qty{0.86}{MA/m} for Co$_{90}$Nb$_5$Zr$_5$ \cite{Zuberek1992}, and \qty{0.88}{MA/m} for Co$_{85}$Nb$_{10}$Zr$_5$ \cite{Sakakima1983}, while values as high as \qty{1.11}{MA/m} have been reported for Co$_{88}$Nb$_8$Zr$_4$ \cite{Takahashi1989} and Co$_{90}$Zr$_{10}$ \cite{Shimada1982}. It is noteworthy that a higher concentration of Co generally results in a higher $M_\mathrm{s}$, as exemplified by $M_\mathrm{s} \approx$ \qty{1.19}{MA/m} for Co$_{96}$Zr$_4$ \cite{Shimada1982}. Conversely, increasing the Zr content tends to reduce $M_\mathrm{s}$ \cite{Ohnuma1980}, as does the addition of Nb \cite{Lekdadri2021}.

To investigate how the inclusion of Nb and Zr reduces the alloy’s magnetism, the saturation magnetic moment per Co atom, $n_\mathrm{B}$, can be estimated using the equation $n_\mathrm{B} = p_\mathrm{Co} M_\mathrm{s}/(\mu_\mathrm{B} N_\mathrm{Co})$,
where $N_\mathrm{Co} = \rho_\mathrm{Co} N_\mathrm{A}/M_\mathrm{Co} \approx \qty{9.09e-22}{cm^{-3}}$ is the Co atom density, derived from the mass density $\rho_\mathrm{Co}$, molar mass $M_\mathrm{Co}$, and Avogadro's constant $N_\mathrm{A}$. The results are summarised in Tab. \ref{Ms_comparison}. For all samples, except pure Co, the estimated magnetic moment per Co atom is significantly lower than expected. This is based on the assumption that the saturation magnetic moment per Co atom increases linearly with the Co content in the alloy, reaching a value of $n_\mathrm{B}$ = \qty{1.71}{\bohrmagneton/\Coatom} at \qty{100}{\percent} Co content, with a saturation magnetisation $M_\mathrm{s}$ = \qty{1.44}{MA/m} \cite{Sun2005}. The observed reduction in $n_\mathrm{B}$ is primarily attributed to the lack of long-range order, which weakens the exchange interactions between Co atoms, thereby reducing the overall magnetisation.

\begin{figure}[htbp]
\centering
\includegraphics[width=0.7\linewidth]{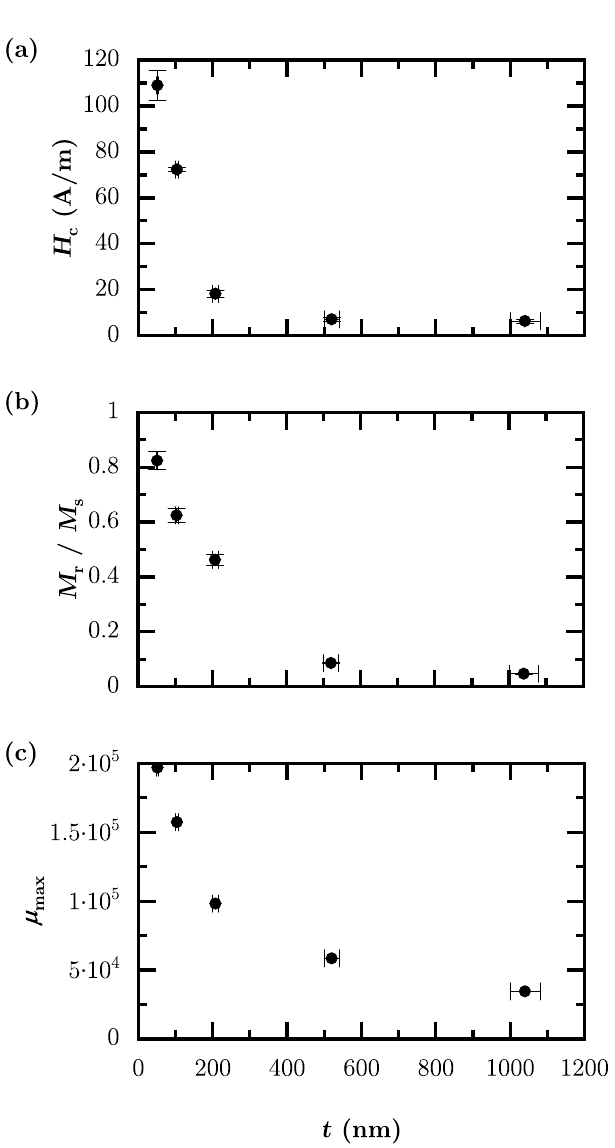} 
\caption{Coercive field $H_\mathrm{c}$, remanent-to-saturation magnetisation ratio $M_\mathrm{r} / M_\mathrm{s}$ and maximum permeability number $\mu_\mathrm{max}$ extracted from the in-plane hysteresis loops of amorphous CoNbZr thin films as a function of film thickness. As the film thickness increases, all the following quantities decrease: (a) $H_\mathrm{c}$ from \qty{110}{A/m} to \qty{10}{A/m}, (b) $M_\mathrm{r} / M_\mathrm{s}$ from 0.82 to 0.05, and (c) $\mu_\mathrm{max}$ from \num{2e5} to \num{4e4}.}
\label{fig:Hys_param}
\end{figure}

Figure \ref{fig:Hys_para} shows the in-plane hysteresis loops of the magnetisation $M$ recorded at a field sweep of $B = \pm$\qty{1}{mT} for amorphous CoNbZr films of varying thicknesses. With increasing thickness, the hysteresis loops evolve from a square to a smoother shape, indicating a change in magnetic behaviour. In addition, Barkhausen jumps are observed in the thinnest film.

From the hysteresis loops in Fig. \ref{fig:Hys_para}, the coercive field $H_\mathrm{c} = B_\mathrm{c}/\mu_0$, the remanent-to-saturation magnetisation ratio $M_\mathrm{r} / M_\mathrm{s}$, and the maximum permeability number $\mu_\mathrm{max}$ were extracted as functions of film thickness, and are presented in Fig. \ref{fig:Hys_param}. As the film thickness increases, all of the following quantities decreases: $H_\mathrm{c}$ from \qty{110}{A/m} to \qty{10}{A/m}, $M_\mathrm{r} / M_\mathrm{s}$ from 0.82 to 0.05, and $\mu_\mathrm{max}$ from \num{2e5} to \num{4e4}. The error bars for $H_\mathrm{c}$ and $M_\mathrm{r} / M_\mathrm{s}$ result from the inaccuracies in the interpolation of the magnetisation between field steps.

A notable observation in Figs. \ref{fig:Hys_param}(a) and \ref{fig:Hys_param}(b) is that the uncertainty in the measure fields decrease with increasing thickness. The increase in film thickness improves the signal-to-noise ratio during the VSM measurements, leading to more precise determinations of $H_\mathrm{c}$ and $M_\mathrm{r}$. The largest error bars at \qty{52}{nm} can be attributed to the Barkhausen jumps, which introduce additional measurement uncertainties. The uncertainty in film thickness increases with thickness for Fig. \ref{fig:Hys_param} due to the constant deposition rate of \qty{4.9(2)}{nm/min}. Longer deposition times required for thicker films amplify the impact of the deposition rate variability, leading to higher absolute uncertainties. 

These findings corroborate previous studies that have also observed a decrease in coercive field with increasing film thickness \cite{Li2007, Cao2012, Takahashi1989}. Additionally, the coercive field at a thickness of \qty{1040}{nm} is consistent with values reported by Shimada et al. (\qty{4}{A/m} \cite{Shimada1982}) and Li et al. (\qty{16}{A/m} \cite{Li2007}).

\begin{figure}[!b]
\centering
\includegraphics[width=\linewidth]{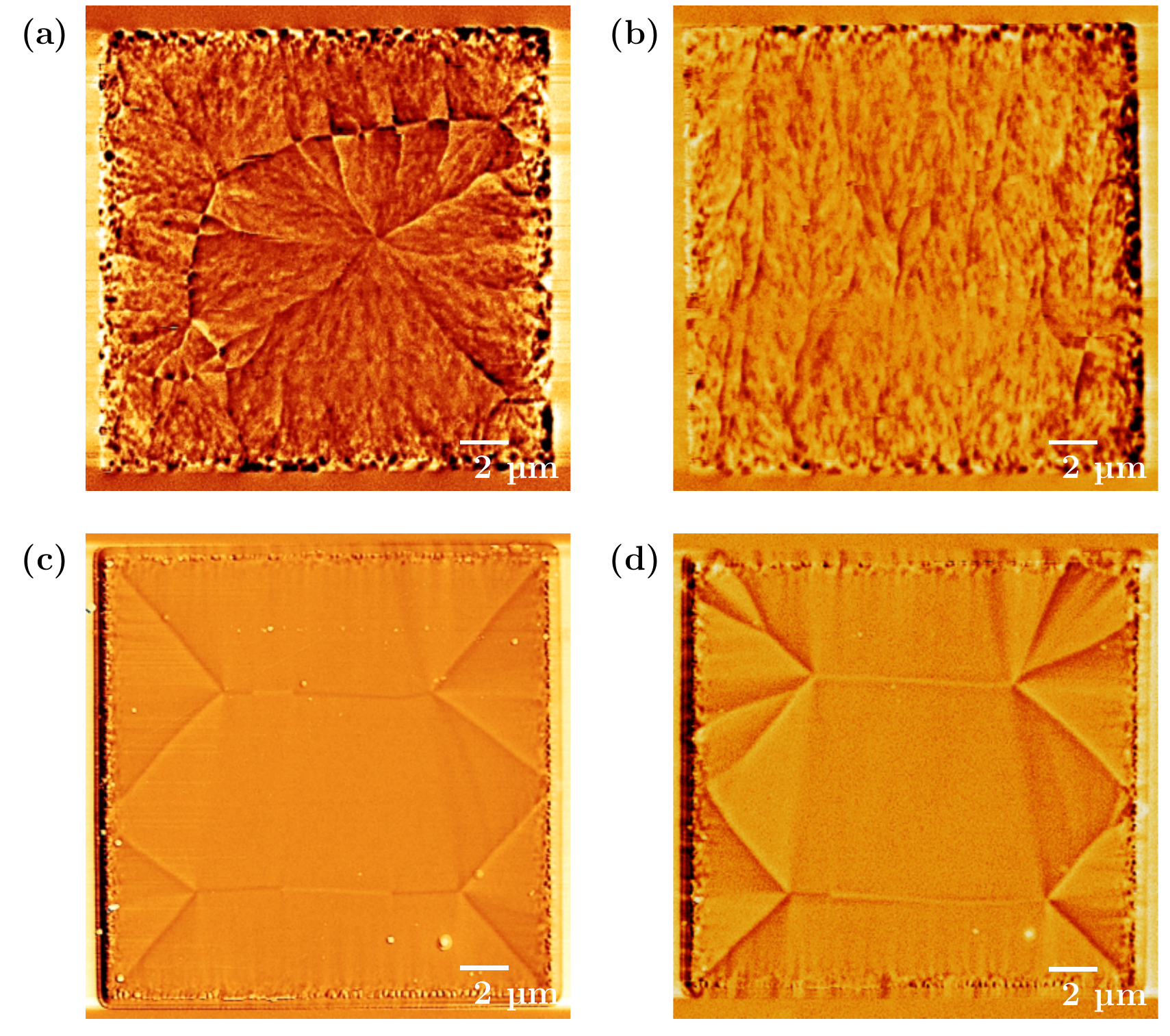} 
\caption{MFM images of \qtyproduct{20x20}{\micro m} patterned CoNbZr films for \qty{52}{nm} (a, b) and \qty{208}{nm} thickness (c, d). Hereby show (a, c) their as-prepared state and (b, d) their remanent state. The images reveal an irregular domain pattern with cross-tie walls at a thickness of \qty{52}{nm}, while a regular flux-closure diamond state is observed at \qty{208}{nm} in both the as-prepared and remanent states.
}
\label{fig:CoNbZr_MFM}
\end{figure}

To explore the remanent magnetisation state, MFM imaging was performed on patterned CoNbZr films with dimensions of \qtyproduct{20x20}{\um} and thicknesses of \qty{52}{nm} and \qty{208}{nm}, as illustrated in Fig. \ref{fig:CoNbZr_MFM}. Hereby show Figs. \ref{fig:CoNbZr_MFM}(a) and \ref{fig:CoNbZr_MFM}(c) their as-prepared state. At a thickness of \qty{52}{nm}, an irregular domain pattern with cross-tie walls is observed, whereas at \qty{208}{nm}, a regular flux-closure diamond state is present. To achieve the remanent state, the samples were saturated with a maximum applied field of \qty{80}{mT}. The resulting remanent states are shown in Figs. \ref{fig:CoNbZr_MFM}(b) and \ref{fig:CoNbZr_MFM}(d). At \qty{52}{nm} thickness, the domains remain irregular, while at \qty{208}{nm}, the domain pattern becomes slightly less regular compared to the as-prepared state, with tulip structures forming at the edges of the elements.

The difference in domain patterns is attributed to the properties of the domain walls themselves. In thinner films, magnetisation tends to remain in-plane, leading to the formation of N\'eel walls, while in thicker films, the magnetisation rotates out of the plane at the centre of the wall, resulting in Bloch walls. Indeed, Fig. \ref{fig:CoNbZr_MFM}(a) shows cross-ties in the thin film, indicating the presence of N\'eel-type domains. In contrast, the domains in Fig. \ref{fig:CoNbZr_MFM}(c) exhibit a combination of \qty{90}{\degree} asymmetric N\'eel walls and \qty{180}{\degree} asymmetric Bloch walls, as indicated by the unipolar contrast in the image. The transition between N\'eel and Bloch walls occurs at approximately $D = 20\sqrt{A/K_\mathrm{d}}$, where $A$ is the exchange constant and $K_\mathrm{d}$ is the anisotropy constant \cite{Hubert2009}. Using a literature value of \qty{11}{pJ/m} for the exchange constant \cite{Yoshihara1994}, the transition from N\'eel to Bloch walls can be calculated with the measured saturation magnetisation. This transition is expected to occur at approximately \qty{84}{nm}, which is consistent with the differences observed between the MFM images at \qty{52}{nm} and \qty{208}{nm}.

\subsection{Hypotheses}

As evident from the hysteresis loops in Fig. \ref{fig:Hys_para}, the remanent magnetisation state and the magnetisation reversal mechanism are different for varying film thickness. At an applied field of \qty{1}{mT}, the thinner films saturate quicker than the thicker ones. This behaviour can be attributed to the varying in-plane demagnetisation factors, $N_x$. The calculated demagnetisation factors are $N_x \approx$ \num{3e-5} for \qty{52}{nm}, $N_x \approx$ \num{9e-5} for \qty{208}{nm}, and $N_x \approx$ \num{40e-5} for \qty{1040}{nm} \cite{Aharoni1998}. Consequently, the demagnetisation field $\mu_0N_x M_\mathrm{s}$ increases from approximately \qty{0.04}{mT} to \qty{0.51}{mT} with increasing thickness. This rise in the demagnetisation field counteracts the external field, requiring thicker films to experience a higher external field to achieve the same level of magnetisation. The variation in the demagnetisation field also affects the remanent magnetisation $M_\mathrm{r}$, as it favours the realignment of the magnetic domains towards a lower energy state once the external field is removed, thereby reducing the remanent magnetisation.

Additionally, the contrast observed in the domain structure of Fig. \ref{fig:CoNbZr_MFM}(b) suggests the presence of a perpendicular anisotropy contribution. The most probable cause might be due to surface anisotropy, which arises from the interfaces or surfaces of the thin film and is inversely proportional to the film thickness \cite{Hubert2009}. This effect is particularly significant in very thin films, where the surface's influence is more pronounced relative to its volume's properties. This is supported by the data in Fig. \ref{fig:Hys_param}, where the observed decrease in coercive field with increasing film thickness is a consequence of the reduced relative contribution of surface anisotropy. In thin films, surface effects strongly influence the total anisotropy, but as the film thickness increases, the volume anisotropy becomes dominant, reducing the overall impact of surface contributions.

\section{Conclusion}

We systematically investigated as-prepared amorphous CoNbZr thin films deposited on Si-SiO$_2$ wafers, varying in thickness from \qty{52}{nm} to \qty{1040}{nm}.

XRD spectra and TEM diffraction patterns confirmed that the CoNbZr films maintain their amorphous structure across all thicknesses. EDX analyses indicated consistent stoichiometry in the sputtered films, with compositions of Co$_{91 \pm 1}$Nb$_{7 \pm 1}$Zr$_{2\pm 1}$. The saturation magnetisation was found to be constant, with a mean value of \qty{1.01(4)}{MA/m}. Based on this saturation magnetisation and the film composition, we estimated that the magnetic moment per Co atom is approximately \qty{60}{\percent} lower than that of pure cobalt. This reduction is attributed to the amorphous nature of the Co matrix and the incorporation of Nb and Zr.

In contrast to the saturation magnetisation, the coercive field exhibits a significant reduction with increasing film thickness, decreasing by a factor of 10 to a value of \qty{10}{A/m}. Similarly, the ratio of remanent-to-saturation magnetisation decreases by a factor of 16, reaching a minimum value of 0.05. The maximum permeability number also decreases by a factor of 5 down to \num{4e4}.

MFM measurements revealed an irregular domain pattern with cross-tie walls at a thickness of \qty{52}{nm}, while \qty{208}{nm} thickness has shown a regular flux-closure diamond state in both, as-deposited and saturated films. This behaviour is attributed to the transition between N\'eel and Bloch type domain walls, which is expected to occur at approximately \qty{84}{nm}.

These results offer valuable insights for the advancement of magnetic sensors based on the giant magnetoimpedance (GMI) effect. The observed trends indicate that sputtering a multilayer system of CoNbZr, with optimised interlayer and film thicknesses, can significantly enhance both the GMI ratio and sensitivity compared to a single-layer CoNbZr film of \qty{1}{\micro \meter} thickness, where the total magnetic thickness remains equivalent. This suggests that multilayer CoNbZr/interlayer thin films may outperform their single-layer counterparts in sensor applications.
\section*{Acknowledgements}

The project was conducted at Saarland University as part of the BMBF-funded collaborative research initiative "ForMikro-spinGMI".

We would like to thank Dr. Oliver Janka from the Service Center for X-ray Diffraction at Saarland University for his invaluable support in collecting the X-ray diffraction data used in this work. We also thank J\"{o}rg Schmauch from INM Saarbr\"{u}cken for his assistance with the TEM measurements.

Further thanks go to Christoph Pauly from the Chair of Functional Materials for his work on FIB structuring and milling, Carsten Brill from KIST Europe for his assistance with film deposition, and Gregor B\"{u}ttel for his valuable contributions throughout the project.

\section*{Declaration of generative AI and AI-assisted technologies in the writing process}

During the preparation of this work, the authors used ChatGPT to improve the readability, language, grammar, spelling, and style of the manuscript. After using this tool, the authors reviewed and edited the content as needed and take full responsibility for the content of the publication.

%

\end{document}